\documentclass[options]{JHEP3}

\newcommand{\be}{\begin{equation}}
\newcommand{\ee}{\end{equation}}
\newcommand{\bea}{\begin{eqnarray}}
\newcommand{\eea}{\end{eqnarray}}

\newcommand{\bit}{\begin{itemize}}
\newcommand{\eit}{\end{itemize}}
\newcommand{\no}{\noindent}

{\title{Generalized Calibrations \& the Characterization of M2-Brane Backgrounds}

\author{Tasneem Zehra Husain\\
Jefferson Laboratory of Physics, Harvard University, Cambridge MA 02138, and\\
LUMS School of Science \& Engineering,
Sector U, D.H.A, Lahore, Pakistan\\
 E-mail: \email{tasneem@physics.harvard.edu}}


\abstract{As a step towards the classification of supergravity backgrounds with flux, we study the (back-reacted) geometry created by a BPS M2-brane when it wraps a cycle in a Calabi-Yau manifold. If it is to preserve supersymmetry, the membrane background must obey certain conditions. These conditions are expressed as geometrical constraints on differential forms and as such, can be interpreted as calibrations. Knowing the complete set of calibrations is the same as satisfying all conditions needed for supersymmetry preservation. While a purely geometric background is completely specified through standard calibrations, in order to fully describe a flux background, we must also state its generalized calibrations. These can be found by probing the background with BPS branes. The logic is simple. Since a BPS probe is guaranteed to be stable, we require that its volume be given by a calibrated form. This applies equally to both charged and uncharged probes; the former are stabilized by flux whereas the latter achieve stability by minimizing their volumes. Volume-forms of charged probes correspond to generalized calibrations and those of uncharged probes, to standard calibrations. Previously geometries were probed only by uncharged branes. The calibrations that were found had then to be supplemented by additional conditions in order to guarantee supersymmetry preservation in backgrounds with flux. Once the scope is broadened to include charged branes, the 'missing conditions' (generalized calibrations) can also be recovered using the probe analysis. We use this method to classify the backgrounds of wrapped M2-branes embedded in Calabi-Yau manifolds by specifying their calibrations.}

\thispagestyle{empty}
\begin{document}

\tableofcontents

\section{Introduction}

It is a fact, well known and much exploited, that a supersymmetric vacuum for string/M-Theory can be created by compactifying flat space on a special holonomy manifold. Perhaps more to the point, this background continues to be supersymmetric as long as the vacuum is populated with purely gravitational objects. The conditions needed to ensure supersymmetry change upon the introduction of flux. These new conditions, and compactification manifolds that satisfy them, are subjects of active study. Ideally, one would like to not only to tame these beasts but also organize them into species, much as Berger was able to do for the special holonomy manifolds\footnote{The reduced, or special, holonomy manifolds are Riemannian manifolds whose holonomy group $G$ is not merely $SO(n)$, but a subgroup thereof. According to Berger's classification, these are the familiar Calabi-Yau $n$-folds, with $SU(n)$ holonomy in $d=2n$, Hyper-Kahler manifolds with $Sp(n)$ holonomy in $d=4n$, $G_2$-holonomy manifolds in $d=7$ and $Spin(7)$ holonomy manifolds in $d=8$}.

The problem is particularly pressing as it turns out that interesting configurations are hardly ever purely geometric. Flux is an intrinsic part of most realistic backgrounds, for instance those generated by branes. Being charged under a gauge potential, branes act as sources for the corresponding field strength hence supersymmetric backgrounds containing branes can not (at least in general) be expected to simply be compactifications of Minkowski space on special holonomy manifolds.In the early years of studying supersymmetric brane configurations however, this is precisely what was assumed. Since special holonomy manifolds were reasonably well understood, the norm was to proceed with the analysis neglecting the effect of flux on the background geometry. This approximation, while a valid first step, is clearly not the whole story; we can not turn off the charge of a brane any more than we can turn off its mass.

Once the subject had matured sufficiently and we had grasped what happens in geometric backgrounds, it was inevitable that we relax our earlier assumption and take the flux into account. What happens to the background now, we ask\footnote{The subject of flux compactification has generated much interest in recent years. See \cite{grana} for a review}. It is precisely this question we will address here, as we focus on the geometries generated by BPS M2-branes.

More specifically, we will work within the framework of 11-dimensional supergravity and explore the supersymmetric background created by an M2-brane wrapped on a cycle in a Calabi-Yau $n$-fold. Being both charged and massive, the brane exerts what is called a back-reaction deforming the geometry into which it is placed. The Calabi-Yau morphs into a manifold $\cal M$ and a field strength flux sourced by the M2-brane propagates through space-time. Our goal is to arrive at the set of constraints that must be imposed on $\cal M$ to ensure that this modified background preserves supersymmetry.

To accomplish this, we will employ the simple but surprisingly telling tool of probe branes. The idea is similar to that in electrodynamics, where we measured the electric field at a point by placing a test particle there; this particle had no impact of its own and could report on its surroundings without disturbing them. Along the same lines, a probe brane is a ghost object which gives rise to no fields itself but merely senses those that are already present. In other words, a probe brane exerts no back-reaction.

Suppose now that the probe we introduce is supersymmetric. It must then also be stable. Stability, we know, is possible only when a system is in an energy minimizing configuration, so a supersymmetric probe has to be in a minimum energy state. In a purely geometric background, minimum energy is synonymous with minimum volume, so a supersymmetric probe will wrap a cycle whose volume form is minimal and hence closed. In more general backgrounds, we can have another class of stable cycles - those that are protected from decay due to the flux that passes through them. A probe brane wrapping this kind of cycle will be stable and will have the minimum energy in its homology class. The volume form of such a probe however, will not be closed but instead related to the space-time flux, as indeed one would expect.

Drawing upon this logic, we will systematically place all possible supersymmetric probes into the wrapped membrane background under investigation. In each case, the statement of stability of a probe can be recast as a constraint on its volume form, or equivalently, on differential forms that are globally defined on $\cal M$.
Simply collecting these conditions might {\it a priori} not seem like much until you recall that this closely parallels the situation that exists for simpler cases in which flux is absent. Only a handful of explicit metrics are known even on the much studied special holonomy manifolds. An entire industry has been set up around Calabi-Yau manifolds, drawing most of its steam from the surprisingly powerful conditions $dJ = 0$ and $d \Omega = 0$. In this paper, we will write down their analogues for the more general back-reacted backgrounds of BPS M2-branes; we can only hope that these constraints prove to be as potent as their special holonomy counterparts!

\section{Supersymmetric Probes \& Calibrations}

In most discussions about BPS branes, it is not long before the word 'calibration' pops up - and for good reason. Calibrations are differential forms characteristic of a particular background and are distinguished in that they enable us to pick out the minimal, and hence preferred, submanifolds. More precisely, a calibration $\phi_p$ is a $p$-form whose pullback onto {\bf any} $p$-dimensional cycle ${\cal N}_p$ is less than or equal to the volume form of the cycle. A cycle that saturates the above bound has the minimum volume in its homology class and is said to be calibrated. Moreover the exterior derivative of a calibrated form is related to the spacetime flux $\cal F$ in the following manner:
\be
{\cal P} (\phi) |_{{\cal N}_p} \leq {\cal V}_{{\cal N}_p} \;\;\;\;\;\;\;\;\;\;\;\;\; d \phi_p = {\cal F}
\label{cal}
\ee
Calibrations are known as {\it standard} when $F=0$ and {\it generalized} otherwise; as is obvious from (\ref{cal}) standard calibrations are closed forms.

In order to appreciate why calibrations are so useful when talking about BPS branes, consider the following.
A $p$-brane probe placed in a spacetime with metric $g_{\mu \nu}$ and $(p+1)$-form gauge potential ${\cal A}$ can be described by the action
\be
{\cal S} =  \int d^{\small{p+1}}\sigma \sqrt{h}  -  \int \cal{P(A)}
\ee
where $h$ is the determinant of the metric $h_{ij} = \partial_i x^{\mu} \partial_j x^{\nu} g_{\mu \nu}$ induced on the worldvolume and ${\cal P}({\cal A})$ is the pullback of the spacetime gauge potential onto the brane. To simplify matters, we can choose static gauge so that ${\cal P(A)}$ is just the restriction of ${\cal A}$ onto the worldvolume.

BPS branes minimize this action and hence satisfy the condition:
\be
\delta S = 0 \;\;\;\;\;\;\ \Rightarrow \;\;\;\;\;\;\ d{\cal V} - {\cal F} = 0
\label{cal}
\ee
where $\cal V$ is the volume form of the $p$-brane, and ${\cal F} = d {\cal A}$ is the $(p+2)$-form  field strength.  Since this equation lies at the heart of our analysis and will be used repeatedly, let us examine it once in detail, so that we are completely comfortable with its mechanics and can proceed to apply it with aplomb to any configuration we come across.

Whether or not a given probe couples to the space-time gauge potential depends on how this probe is embedded into the background. If the pull back (restriction) of the gauge field onto its worldvolume is non-vanishing, the brane is charged under the potential, i.e a $p$-brane with volume-form ${\cal V} = {\cal V}_{{\mu}_0 \dots {\mu}_p} \; dx^{{\mu}_0} \wedge \dots \wedge  dx^{{\mu}_p}$ carries a charge {\bf only if} the spacetime gauge potential contains a term that goes like ${\cal A}_{{\mu}_0 \dots {\mu}_p} \; dx^{{\mu}_0} \wedge \dots \wedge dx^{{\mu}_p}$.

In order to treat on an equal footing, branes that couple to the gauge field electrically and magnetically, it turns out be more convenient to work with the field strength\footnote{For probes that couple magnetically to the gauge potential, the dual field strength is all we really know}, so practically speaking what we will find ourselves requiring is that
\be
\partial_{\nu} {\cal V}_{{\mu}_0 \dots {\mu}_p} dx^{\nu} \wedge dx^{{\mu}_0} \dots dx^{{\mu}_p} =
{\cal F}_{ \nu {\mu}_0 \dots {\mu}_p} dx^{\nu} \wedge dx^{{\mu}_0} \dots dx^{{\mu}_p}
\ee
should hold independently for each term in $d {\cal V}$. Clearly, those terms that have no counterparts in
${\cal F}$ must vanish.

The similarity of this condition to the definition (\ref{cal}) is striking and not entirely coincidental. In fact, a
stable $p-$brane has the minimum energy of all $p-$branes wrapping cycles in the same homology class and thus wraps a calibrated cycle. The worldvolume of a supersymmetric brane thus corresponds to a calibrated form in the ambient space. As a result, we find that the calibrations of a particular background can be found merely by glancing at the supersymmetric probes it admits. The volume form of a charged probe corresponds to a generalized calibration whereas an uncharged probe, which feels only the geometric aspects of the background has as its volume form a standard calibration. Since the volume of a probe is measured using differential forms, each calibration can be expressed as a condition on these forms that are globally defined and characteristic of the background.

To place this work in context, we should mention that it is the logical culmination of a series of ideas stretching as far back as \cite{gcal}, where the authors used $Mp-$brane probes to investigate the supergravity background of a single planar $Mp-$brane. In \cite{amherst} and \cite{thesis} it was shown how the power of calibration technology could be harnessed to write down the supergravity solutions of wrapped M-branes. Probe branes were first used to investigate such backgrounds in \cite{fh} and these methods were subsequently applied to several wrapped D-brane systems in \cite{examples}.

We now show how this scheme can be taken one step further. The probes considered in \cite{fh} only wrapped minimum volume cycles. This was not without reason; prevailing wisdom \cite{folklore} seemed to suggest that probes which
overlap completely with the background brane should be avoided, as should those which span the entire space transverse to it. The former condition could be satisfied by insisting that all probes have a sensible interpretation in the brane's worldvolume theory; the latter had to be enforced by hand.

However, neither of these requirements is essential. The first can be waived simply by
realizing that there is no reason to appeal to the worldvolume theory for a justification of probes which are perfectly sensible objects with a well defined space-time interpretation. In fact, since we want to study the full 11-dimensional geometry, it actually makes sense to step off the brane and adopt a space-time point of view. This insight allows us to include, as legitimate tools in our analysis, probes that span the entire worldvolume of the background brane\footnote{This is, perhaps, not very relevant for the case at hand, since only the time direction remains 'unwrapped', but it is an important philosophical difference of approach, and one that will play an important role in discussions of wrapped fivebrane backgrounds}.

In order to address the second concern, recall that both standard and generalized calibrations correspond to volume-forms of probes wrapping supersymmetric cycles. In the former case, these cycles are stable because of their minimum volume and in the latter, cycles are stablized by the presence of spacetime flux. Notice in particular, that generalized calibrations correspond to probes that couple to the background gauge field. Depending on whether they are M2 or M5-branes, this means that the probes span either the entire space tangent to the brane, or the entire space transverse to it. This explains why a self-consistent, but incomplete, set of constraints was obtained in \cite{fh}. Probes that couple to the gauge field were explicitly excluded from the analysis, and we found only what we looked for - the standard calibrations.

In this paper, we extend the analysis of \cite{fh} as we place {\it every} possible supersymmetric M-brane probe into the wrapped brane geometries under study. This enables us to arrive at a comprehensive list of constraints\footnote{Several of these were previously known \cite{membranes}, but their origin was somewhat mysterious.} (or calibrations) which can serve to classify these backgrounds.

\section{BPS Membranes \& Calabi-Yau Manifolds}

Armed with these generalities, it is now time to turn to specifics. As mentioned earlier, we will focus on the supergravity background created by an M2-brane as it wraps a supersymmetric cycle in a Calabi-Yau manifold\footnote{Even though we deal explicitly only with Calabi-Yau manifolds in this paper, the analysis presented here generalizes to the other special holonomy manifolds as well}.  The back-reaction of the brane introduces a flux and causes the Calabi-Yau to deform into a manifold $\cal M$. In order to get a handle on $\cal M$, we need to understand how the back-reaction works and the ways in which it modifies space-time.

Recall that the M2-brane is a classical object in supergravity and can hence be turned on slowly and continously. The deformation that the Calabi-Yau $n$-fold undergoes in response to the brane is consequently a smooth process which does not change the topological structure, hence tensors that were globally defined on the Calabi-Yau manifold are 'inherited' by $\cal M$. The Calabi-Yau $n$-fold, and hence $\cal M$, admit a metric $g_{ab}$, a 2-form $J_{ab}$ and a $(n,0)$ volume form $\Omega$; the existence of an almost complex structure ${\cal I}$ is implicit in the above\footnote{The 2-form $J$ is constructed by using the metric to lower an index on the almost complex structure $J_{ab} = {\cal I}_a^{\;c} g_{cb}$, where ${\cal I}_a^{\; c} \; {\cal I}_c^{\; b} = - \delta_a^{\; b}$}. The presence of this almost complex structure is very useful, since it allows us to define the notion of holomorphicity at least locally. We can refer to forms on $\cal M$ as being of type $(p,q)$ as usual, as long as we keep in mind that this classification is with respect to tangent space indices. Notice that $g, J$ and $\Omega$ are invariant under $SU(n)$ transformations of the frame bundle; as such, they are sometimes referred to as (defining) the $SU(n)$ structure on ${\cal M}$.

Despite these similarities in their global structure, ${\cal M}$ looks quite different to the Calabi-Yau locally. This change, which is essentially a measure of the back-reaction of the brane, is manifested in the differential conditions obeyed by the tensors $g, J$ and $\Omega$. On a Calabi-Yau $n$-fold, the constraints $dJ = 0$ and $d \Omega = 0$ hold, reflecting the fact that the manifold is Kahler and has $SU(n)$ holonomy; this will no longer be true on ${\cal M}$. Our aim is to come up with an analogous set of conditions that are valid on, and hence can serve to classify the back-reacted manifold.

Before we proceed any further, a few general remarks about supergravity solutions are perhaps in order\footnote{See \cite{douglas} for a comprehensive review}. Recall that the only fields in 11-dimensional supergravity are a metric, a three-form gauge potential $\cal A$ and the gravitino $\Psi$. Since we are interested in bosonic supersymmetric solutions, we set the gravitino to zero which results in the supersymmetry transformations of the bosonic fields vanishing automatically. All that is required now for supersymmetry is that the variation of the gravitino $\delta_{\epsilon} \Psi$ vanish as well. Einstein's equations are guaranteed to be satisfied if, in addition, the field strength obeys the Bianchi Identity and equations of motion\footnote{Strictly speaking, there should also be source terms, indicating the position of the membrane}. A bosonic supersymmetric solution thus consists simply of a metric and field strength such that  $\delta \Psi = 0$, ${\cal F} =  0$ and $d {\ast} {\cal F} = 0$. Satisfying these constraints turns out to be much simpler than solving Einstein's second order differential equations, so one way to proceed is the following\cite{douglas}, \cite{fs}:

Based on the isometries of a particular BPS configuration, we can write down the general form of the metric. The field strength follows simply from the fact that the volume form of a BPS brane is (gauge) equivalent to the electric potential under which the brane is charged\footnote{An easy way to see this is to realize that the BPS condition equates the mass of the brane with its charge and that the charge and mass are given essentially by the integrals of the gauge potential and volume-form respectively}. These ansatze for the metric and field strength are then plugged into the equation $\delta_{\epsilon} \Psi = 0$. The relations that result from requiring this equation to hold allow us to express all the unknowns in our ansatze in terms of a single quantity, and in addition, impose certain constraints on the background. Of the Bianchi Identity and equations of motion, one will be satisfied automatically (which one it is depends on whether the background is generated by an M2 or an M5-brane) and the other takes the form of a non-linear differential equation. Solving this equation determines the one remaining unknown and hence leads to a complete supergravity solution. In practice, this equation has proved difficult to solve, except in the near horizon limit. Notice, however, that the constraints on the background come about simply from requiring supersymmetry. These had been  painstakingly computed in earlier work, until we found in \cite{fh} that the entire set of these previously known constraints and then some, can be reproduced simply by placing probes in the background, which is the approach we will take here.

Since we want to create a BPS background by embedding an M2-brane inside a Calabi-Yau manifold, we must ensure that the brane wraps a supersymmetric cycle. Calabi-Yau manifolds admit two types of supersymmetric cycles - even dimensional holomorphic $r$-cycles calibrated by $J^r$ (for all $r  \leq n$) and a unique $n$-dimensional Special Lagrangian (SpelL) $n$-cycle calibrated by $Re (\Omega e^{i \theta})$. Supersymmetric configurations can thus be created either by (a) wrapping both spatial directions of the membrane on a holomorphic two-cycle (which is possible in any Calabi-Yau manifold) or (b) wrapping the entire worldvolume of a Euclidean membrane on a SpelL 3-cycle in a Calabi-Yau three-fold.

 We want to employ as probes only those branes that will preserve (some of) the supersymmetries of the background so let us briefly remind ourselves when this is possible. Two intersecting membranes form a supersymmetric configuration if they share one space-time direction, whereas a membrane-fivebrane system preserves supersymmetry {\it either} when the membrane ends on the fivebrane {i.e shares two space-time directions} {\it or} when the two are completely transverse to each other. In addition to intersecting the background brane in the appropriate way, an acceptable probe must also wrap a supersymmetric cycle in the compactification manifold. Since ${\cal M}$ inherits the tensors $J$ and $\Omega$, supersymmetric cycles in $\cal M$ continue to be calibrated by $J^r$ and $Re (\Omega e^{i \theta})$.

In order to keep track of the probes and to figure out the cycles they can be wrapped on, it is convenient to employ what is called a brane scan; in this tabular representation of the brane configuration, we will denote by $\otimes$ directions spanned by the M2-brane that gives rise to the background and use $\times$ to indicate directions that are along the worldvolume of a probe. Also, to avoid confusion (and unnecessarily wordy sentences) we will from now on refer to probe branes as membranes or fivebranes, reserving the term 'M2-brane' exclusively for the membrane that gives rise to the background.

\section{M2-branes on Holomorphic Cycles}

Consider the background of an $M2$-brane wrapped on a holomorphic cycle in the $2n$-dimensional manifold ${\cal M}_n$. Due to the presence of the brane space-time is 'divided' into three distinct subspaces: the time direction, the manifold ${\cal M}_n$ (which contains the spatial part of the M2-brane) and the $10 - 2n$ directions that are left over.

In what follows, the time direction is denoted interchangeably by $t$ and $x^0$. To facilitate comparison between configurations where the dimension $n$ differs, we will not
represent the 10 spatial directions as $x^i$, but instead adopt a convention that makes apparent the split between the compactification manifold and the rest of space. Directions transverse to ${\cal M}_n$ are denoted by $y^a = x^{2n +a}$, $a = 1 \dots 10 - 2n$, while ${\cal M}_n$ itself is spanned by coordinates $z^m = x^m + ix^{m+n}$ for $m = 1 \dots n$. To avoid a proliferation of indices, we will denote by $\partial$, $\bar{\partial}$ and $\partial^{'}$ the derivatives with respect to $z^m$, $\bar{z}^n$ and $y^a$, i.e
\be
df = \partial f + {\bar \partial} f + \partial^{'}f, \;\;\;\;\;\;\;\;\;\;
\partial f = (\frac{\partial f}{ \partial z^m}) dz^m, \;\;\;\;
{\bar \partial} f = (\frac{\partial f}{ \partial {\bar z}^n}) d{\bar z}^n, \;\;\;\;
\partial^{'}f = (\frac{\partial f}{ \partial y^a}) dy^a
\ee

Since ${\cal M}_n$ inherits an almost complex structure $\cal I$ from the underlying Calabi-Yau, we are allowed to define coordinates $z^m$. However, we do not know a priori if $\cal I$ is integrable, so for now we will just assume that these coordinates are locally holomorphic. While our entire analysis could equally well have been carried out in real coordinates, we exercise our right to define $z^m$ because the expressions for the metric, $J$ and $\Omega$ are simpler in complex coordinates. This decision is vindicated later when we see that for all geometries created by an M2-brane wrapping a cycle calibrated by $J$, (a suitably rescaled version of) the volume form $\Omega$ is closed so a complex structure exists on ${\cal M}_n$ and holomorphic coordinates can be defined.

With our notation fixed, we can now attempt to describe the supergravity background of the M2-brane. To do so, we need expressions for the metric and the four-form field strength.A glance at the isometries of the configurations tells us that physical quantities should be invariant under rotations in the transverse directions so the metric must be of the form
\be
ds_{11}^2 =  - H_1^{2} dt^2 + 2 G_{m {\bar n}} dz^m dz^{\bar n} + H_2^{2} \delta_{a b} dy^{a} dy^{b}
\label{metric}
\ee
 This has come to be known as the Fayyazuddin-Smith metric ansatz. No assumption is made about the functions $H_1$ and $H_2$ or the metric $G_{m {\bar n}}$ except that they are independent of time. The expression for the four-form can be obtained by recalling that the electric potential $\cal A$ to which the brane couples is gauge equivalent to its volume form $\cal V$. A membrane wrapping a holomorphic cycle in the above space-time has a volume form
\be
{\cal V} = H_1 dt \wedge J
\ee
where $J_{m{\bar n}} = i G_{m {\bar n}}$. Consequently, the field strength ${\cal F} = d{\cal A}$ has components
\bea
{\cal F}_{0mn{\bar p}} &=& \frac{i}{2} [\partial_{n} (H_1 G_{m {\bar p}}) - \partial_{m} (H_1 G_{n {\bar p}})] \nonumber \\
{\cal F}_{0mna} &=& - i \partial_a (H_1 G_{m {\bar n}})
\label{f}
\eea
It follows that that $d {\cal F} = 0$ will automatically be satisfied. We will, however, still need to impose the condition $d \ast {\cal F} = 0$.

While the general structure of the background is reflected in eqns $(\ref{metric}), (\ref{f})$, the functions remain unknown. We will determine these functions not by imposing $\delta \Psi = 0$ but rather by probing the background using supersymmetric branes. For membranes, the story is quite simple; only a probe which lies entirely parallel to the background brane can couple to ${\cal A}$. Even though our current formalism can accomodate this probe, we will not mention it in future discussions, since it yields no new information\footnote{Since the probe has a volume form identical to that of the background brane it just reiterates (\ref{f})}. All other supersymmetric membranes are blind to ${\cal F}$ and must therefore have closed volume forms.

Notice that up to now, we have had no reason to specify the dimension $n$ of the manifold ${\cal M}_n$ - our analysis so far applies equally to them all. Now, however, we have reached a branching point. The components of the dual seven-form $\tilde{\cal F} =  \ast {\cal F}$ depend on $n$ and hence will be felt by different fivebrane probes in each case. So, unlike for the membranes, we can not make a sweeping statement about the volume forms of fivebrane probes; these will simply need to be analyzed one by one and we will have to see whether for a given probe, any component in $d {\cal V}$ has a counterpart in $\tilde{\cal F}$.

Having built up all the necessary machinery, we are now in a position to explore the backgrounds created by membranes wrapping holomorphic curves in Calabi-Yau $n$-folds for $n = 2 \dots 5$. In addition to reproducing the constraints on $\partial J^{n-1}$, we show that (upto a warp factor) $\Omega$ is always closed. We are also able to obtain constraints on $\partial^{'} J^n$, the transverse space derivative of the volume form on ${\cal M}_n$. To the best of our knowledge, these constraints have not previously been explicitly written down.

\subsection{M2 on a holomorphic cycle in ${\cal M}_2$}

The manifold ${\cal M}_2$ is spanned by $z^1, z^2$ and there are six transverse directions $y^1 \dots y^6$.
Our M2-brane is wrapped on the 2-cycle calibrated by $J$, as represented below\footnote{Since $J = dz^1 \wedge d{\bar z}^1 + dz^2 \wedge d{\bar z}^2$, the volume form of the M2-brane is equal to the pullback of $J$ onto it}.
\be
\begin{array}[h]{|c|c|cccc|cccccc|}
  \hline
   \; & 0 & 1 & 2 & 3 &
              4 & 5 & 6 & 7 & 8 & 9 & 10\\
  \hline
  {\bf M2} & \otimes & \otimes & & \otimes & & & & & &  &  \\
  \hline
\end{array}
\ee
The metric and field strength are, as always, given by (\ref{metric}) and (\ref{f}). Branes that can be brought into this background without entirely breaking supersymmetry are acceptable probes. We list these below:
\be
\begin{array}[h]{|c|c|cccc|cccccc|}
  \hline
   \; & 0 & 1 & 2 & 3 &
              4 & 5 & 6 & 7 & 8 & 9 & 10\\
  \hline
   {\bf M2} & \otimes & \otimes & & \otimes & & & & & &  &  \\
   \hline
   (a) \; {\it M2} & \times & & & & & \times & \times  &  &  &  &   \\
   (b) \; {\it M2} & & \times &  \times & & & \times &  &  &  &  &   \\
   (c) \; {\it M5} & \times & \times &  \times& & & \times & \times  & \times &  &  &   \\
   (d) \; {\it M5} &  & \times & \times & \times & \times & \times &  \times &  &  &  &   \\
   (e) \; {\it M5} & & \times & & \times & & \times & \times  & \times  & \times &  &   \\
   (f) \;  {\it M5} &  &  & &  & & \times & \times & \times & \times & \times & \times  \\

  \hline
\end{array}
\ee
Recall that the membranes always have closed volume forms, so for both (a) and (b) we simply need to impose $d {\cal V} = 0$.
Fivebranes need a little more work. Given the structure of space-time and the expression for the four form (\ref{f}), it follows that the dual seven-form $\tilde{\cal F}$ has components $\tilde{\cal F}_{mabcdef} = \ast {\cal F}_{0 {\bar m} {\bar n} p}$, its complex conjugate $\tilde{\cal F}_{{\bar m} abcdef} = \ast {\cal F}_{0 m n {\bar p}}$ and $\tilde{\cal F}_{m {\bar n} abcde} = \ast {\cal F}_{0 p {\bar q} f}$. From the table above, it is easy to see that only probes (e) and (f) can couple to this field strength . For the former, $d{\cal V}$ has contributions from ${(\partial {\cal V})}_{mabcdef}$ and ${(\bar{\partial} {\cal V})}_{{\bar m} abcdef}$ whereas for the latter there is a term ${(\partial^{'} {\cal V})}_{m{\bar n} abcde}$. The remaining two fivebrane probes (c) and (d) are uncharged and hence will have closed volume forms.

We now write down the constraints that correspond to from each probe:
\bit
\item
(a) For this first membrane probe, ${\cal V}= H_1 H_2^2 dt \wedge dy^1 \wedge dy^2$, so $d {\cal V} = 0$ implies
\bea
\partial (H_1 H_2^2) \wedge dt \wedge dy^1 \wedge dy^2 = 0 \nonumber \\
\bar{\partial} (H_1 H_2^2) \wedge dt \wedge dy^1 \wedge dy^2 = 0 \nonumber \\
\partial^{'} (H_1 H_2^2) \wedge dt \wedge dy^1 \wedge dy^2 = 0 \nonumber
\eea
All these conditions are satisfied if $H_1 H_2^{2} = f (y^1, y^2)$. Notice, however, that we could have picked the M2-probe to span {\it any} two of the six transverse directions $y^1 \dots y^6$. Each such choice leads to a condition like the one above, and we end up with the requirement that
$$H_1 H_2^{-2} = f (y^1, y^2) = f(y^1, y^3) = .... = f(y^5, y^6)$$
which is satisfied only if the function $f$ is a constant. We pick this constant to be 1, for convenience, and let
\be
H_1 = H_2^{-2} \equiv H^{-1/3}
\label{h1h2}
\ee
Setting $H_1 =  H^{-1/3}$ is a nod to convention. In doing so, we are drawing an analogy with the solution for a single planar membrane $ds^2 = H^{-2/3} dx^2_{||} + H^{1/3} dx^2_{\perp}$ where $x_{||}$ and $x_{\perp}$ represent coordinates tangent and transverse to the brane respectively, and $H = H(x_{\perp})$ is a harmonic function.
\item
\no
(b) The volume form of this Euclidean membrane is ${\cal V}= H_2 \Omega \wedge dy^1$ so $d {\cal V} = 0$ implies
$$\partial (H_2 \Omega) \wedge dy^1 = 0, \;\;\;\;\;\;
\bar{\partial} (H_2 \Omega) \wedge dy^1 = 0, \;\;\;\;\;\;
\partial (H_2 \Omega) \wedge dy^1 = 0$$
These conditions are satisfied if $H_2 \Omega = g(y^1)$. Again, as for the previous probe, this membrane could equally well span any one of the six transverse coordinates, so what we actually end up with is the statement
$$H_2 \Omega = g(y^1) \dots = g(y^6)$$which is satisfied only when the function $g$ is a constant, i.e
\be
d (H_2 \Omega) = 0
\label{omeg}
\ee
\item
(c) For the fivebrane wrapped on the SpelL 2-cycle, ${\cal V}= H_1 H_2^3 dt \wedge \Omega \wedge dy^1\wedge dy^2 \wedge dy^3$ is closed. The resulting constraints
$$\partial (H_1 H_2^3 \Omega) = 0, \;\;\;\;\;\;\;\;\;\;\;\; \bar{\partial} (H_1 H_2^3 \Omega) = 0, \;\;\;\;\;\;\;\;\;\;\;\; {\partial}^{'} (H_1 H_2^3 \Omega) = 0$$
are merely a combination of (\ref{h1h2}) and (\ref{omeg}) and yield no new information.
\item
(d) For this fivebrane probe too, ${\cal V} = H_2^2 J \wedge J \wedge dy^1 \wedge dy^2$ is closed. Since ${\cal V}$ saturates all the indices along ${\cal M}_2$, $\partial {\cal V}$ vanishes identically. The only constraint we obtain comes from requiring that the transverse space derivative vanish also, i.e
\be
\partial^{'} (H_2^2 J \wedge J) = 0
\label{cy2}
\ee
\item
(e) The Euclidean five-brane has volume form${\cal V} = H_2^4 J \wedge dy^1 \wedge dy^2 \wedge dy^3 \wedge dy^4$. Of the terms in $d{\cal V}$, neither $\partial {\cal V}$ nor its complex conjugate $\bar{\partial} {\cal V}$ have a counterpart in $\tilde{ \cal F}$ and hence vanish, implying that
\be
\partial (H_2^4 J) = {\bar \partial} (H_2^4 J) = 0
\label{j2}
\ee
The term $\partial^{'}{\cal V}$, however, has an index structure which matches that of $\tilde{ \cal F}_{m{\bar n} abcde}$, so the two quantities should be equal:
\be
\partial^{'} (H_2^4 J)  \wedge dy^a \wedge dy^b \wedge dy^c \wedge dy^d = \ast \partial^{'} (H_1 J) \wedge dt
\label{fa}
\ee
This can be used to show that
\be
\partial^{'} \ln \sqrt{G} = - \frac{1}{3} \partial^{'} \ln H
\label{h'}
\ee
where $G$ is the determinant of the metric $G_{m {\bar n}}$.
\item
(f) The volume form of this Euclidean fivebrane ${\cal V}_{abcdef} = H_2^6 dy^1 \wedge \dots \wedge dy^6$ saturates all transverse space indices, hence $\partial^{'} {\cal V}$ vanishes automatically. The remaining two derivatives $\partial {\cal V}$ and $\bar{\partial} {\cal V}$ have just the right index structure to balance ${\tilde {\cal F}}_{m123456}$ and ${\tilde {\cal F}}_{{\bar m} 123346}$ respectively. Writing these out, we have
\be
\partial H_2^6 \wedge dy^1 \wedge \dots \wedge dy^6 = \ast \partial (H_1 J) \wedge dt
\label{fb}
\ee
together with its complex conjugate. These imply the constraints
\be
\partial \ln {\sqrt G} = - \frac{1}{3}\partial \ln H, \;\;\;\;\;\;\;\;\; {\bar \partial} \ln {\sqrt G} = - \frac{1}{3} {\bar \partial} \ln H
\ee
\eit

\no
To summarize, then, we have found that the bosonic background of a membrane wrapping a holomorphic curve in ${\cal M}_2$ is given by the metric and field-strength
\bea
ds_{11}^2 &=&  - H^{-2/3} dt^2 + 2 G_{m {\bar n}} dz^m dz^{\bar n} + H^{1/3} \delta_{a b} dy^{a} dy^{b} \\
{\cal F}_{0mn{\bar p}} &=& \frac{i}{2} [\partial_{n} (H^{-1/3} G_{m {\bar p}}) - \partial_{m} (H^{-1/3} G_{n {\bar p}})] \nonumber \\
{\cal F}_{0m{\bar n}a} &=& - i \partial_a (H^{-1/3} G_{m {\bar n}})
\label{mfs}
\eea
The determinant $G$ of the metric $G_{m {\bar n}}$ is related to $H$ through $\sqrt{G} = H^{-1/3}$. In addition, ${\cal M}_2$ is constrained by supersymmetry to be such that the following relations hold, together with their complex conjugates, where applicable.
\bea
\begin{array}{ccc}
\bar{\partial} (H^{1/6} \Omega) = 0 \;\;\;\;\;\;\; \partial^{'} (H^{1/6} \Omega) = 0 \nonumber  \\
\partial (H^{2/3} J) = 0  \nonumber  \\
\partial^{'} (H^{1/3} J \wedge J) = 0
\label{m2cond}
\end{array}
\eea
Notice that $H^{1/6} \Omega$ is a closed form in spacetime but $H^{2/3} J$ is closed only on ${\cal M}_2$, since $\partial^{'} (H^{2/3} \Omega) \neq 0$. To guarantee that Einstein's equations are solved, the above conditions (\ref{m2cond})should be supplemented by the equation of motion for the field strength $d \tilde{\cal F}$. Reading off the components of $\tilde{\cal F}$ from (\ref{fa}, \ref{fb}) this equation becomes:
\be
 \partial^{'} \partial^{'} (H^{2/3} J) + 2 i \partial \bar{\partial} H = 0
\ee

\subsection{M2 on a Holomorphic Cycle in ${\cal M}_3$}

Suppose now that the M2-brane wraps a holomorphic two-cycle embedded in the six-dimensional manifold ${\cal M}_3$. Coordinates $z^m (m = 1, 2 \dots 3)$ span ${\cal M}_3$ and transverse directions are represented by $y^a (a = 1, 2 \dots 4)$. Since the metric and four-form are once again given by (\ref{metric}) and (\ref{f}) respectively, the dual seven-form has components ${\tilde {\cal F}}_{mn{\bar p} abcd}$, ${\tilde {\cal F}}_{m {\bar n} {\bar p} abcd}$ and ${\tilde {\cal F}}_{mn {\bar p} {\bar q} abc}$.

We will now list the branes that can be brought into this background without entirely breaking supersymmetry, and write down the conditions that follow from using these as probes.
\be
\begin{array}[h]{|c|c|cccccc|cccc|}
  \hline
   \; & 0 & 1 & 2 & 3 &
              4 & 5 & 6 & 7 & 8 & 9 & 10\\
              \hline
  {\bf M2} & \otimes & \otimes & & & \otimes  & & & & &  &  \\
  \hline
  (a) \; {\it M2} & \times & & &  & & & & \times & \times &  &  \\
  (b) \; {\it M2} &  & \times & \times & \times & & & &  &  &  &  \\
 (c) \; {\it M5} & \times & \times & \times & \times & & & & \times & \times &  &  \\
 (d) \; {\it M5} & & \times & \times & \times & \times  & \times & \times & & & & \\
 (e) \; {\it M5} & & \times & & & \times  & & & \times & \times & \times & \times \\
(f) \; {\it M5} &  & \times & \times & & \times & \times & & \times & \times &  &  \\
   \hline
\end{array}
\ee
Of these probes, the only ones that will feel the field strength are (e) and (f); the rest all have closed volume forms.
\bit
\item
(a) The volume-form ${\cal V} = H_1 H_2^2 dt \wedge dy^1 \wedge dy^2$ of the membrane is closed, i.e,
\be
d (H_1 H_2^2) \wedge dt \wedge dy^1 \wedge dy^2 = 0
\ee
Arguing as we did earlier, we find $H_2^{-2} = H_1$ just as in (\ref{h1h2}). Once again, we set
\be
H_1 = H_2^{-2} \equiv H^{-1/3}
\ee
\item
(b) Again, as in ${\cal M}_2$, for the membrane probe wrapping the SpelL cycle we have a choice to make. Depending on the phase we pick, we can wrap the probe on the cycle calibrated by $Re \Omega$ or $Im \Omega$, both are supersymmetric\footnote{Of course we could equally well have chosen cycles inbetween, but these two will suffice to illustrate the point we are trying to make.} In either case, the volume form is closed and we obtain the constraints $d (Re \Omega) = 0$ and $d (Im \Omega) = 0$. Combining these we reach the conclusion that
\be
d \Omega = 0
\ee
\item
(c) The fivebrane probe wrapped on a SpelL 3-cycle has a closed volume-form, so we find
\be
d (H_1 H_2^2 \; \Omega) \wedge dt \wedge dy^a \wedge dy^b = 0
\ee
which is a reiteration of previously known constraints.
\item
(d) This fivebrane completely wraps ${\cal M}_3$, so derivatives $\partial{\cal V}$ and ${\bar \partial} {\cal V}$ vanish trivially. The only relevant contribution to $d{\cal V}$ comes from $\partial^{'} {\cal V}$. The volume form ${\cal V} = J \wedge J \wedge J$ is closed only if
\be
\partial^{'} (J \wedge J \wedge J) = 0
\ee
\item
(e) For the Euclidean fivebrane on the 2-cycle, ${\cal V} = H_2^4 J \wedge dy^{1} \wedge dy^{2} \wedge dy^{3} \wedge dy^{4}$. Since the brane wraps all transverse directions, the derivative $\partial^{'} {\cal V}$ vanishes automatically due to antisymmetry and this probe can not be used to draw any conclusions about $\partial^{'} (H_2^4 J)$. The component $\partial {\cal V}$ on the other hand, has the same index structure as ${\tilde {\cal F}}_{mn{\bar p}abcd}$ implying the following constraint
\be
\partial (H_2^4 J) \wedge dy^1 \wedge dy^2 \wedge dy^3 \wedge dy^4 = \ast \partial (H_1 J) \wedge dt
\ee
which can be used to show that $\partial \ln \sqrt{G} = 0$.
\item
(f) Our final probe is the fivebrane with worldvolume ${\cal V} = H^2_2 J \wedge J \wedge dy^{a} \wedge dy^{b}$. The field strength ${\tilde {\cal F}}$ does not contain any term with an index structure to match that of $\partial {\cal V}$, so
$$\partial (H_2^2 J \wedge J) = 0$$
The component ${\tilde {\cal F}}_{mn{\bar p} {\bar q} abc}$, however, has the same index structure as $\partial^{'} {\cal V}$ so the two quantities must be equal, i.e
$$\partial^{'} (H_2^2 J \wedge J) \wedge dy^a \wedge dy^b = \ast \partial^{'} (H_1 J) \wedge dt$$
This can be used to show that $\partial^{'} \ln \sqrt{G} = 0$.
\eit

\no
Collecting these results we find that the background of an M2-brane wrapping a holomorphic 2-cycle in ${\cal M}_3$ can be described by the metric (\ref{metric}) and field strength (\ref{f}) with the determinant $G$ now being constant. The compactification manifold ${\cal M}_3$ is constrained by supersymmetry to obey
\bea
d \Omega = 0 \nonumber  \\
\partial (H^{1/3} J \wedge J) = 0 \nonumber  \\
\partial^{'} (J \wedge J \wedge J) = 0 \nonumber
\eea
The equation of motion for the field strength can be expressed as follows:
\be
\partial^{'} \partial^{'}(H^{1/3} J \wedge J) + 4 i \partial \bar{\partial} (H^{2/3} J) = 0
\ee

\subsection{M2 on a Holomorphic Cycle in ${\cal M}_4$}

When the two-cycle wrapped by the M2-brane is embedded into an eight-dimensional manifold ${\cal M}_4$, the metric and field strength are once again of the form (\ref{metric}) and (\ref{f}) except that now there are four coordinates $z^m, m = 1 \dots 4$ and only two transverse coordinates $y^a, a=1,2$.
\no
The supersymmetric probes of this geometry are shown below:
\be
\begin{array}[h]{|c|c|cccccccc|cc|}
  \hline
   \; & 0 & 1 & 2 & 3 &
              4 & 5 & 6 & 7 & 8 & 9 & 10\\
  \hline
  {\bf M2} & \otimes & \otimes &  & & & \otimes& & & &  &  \\
  \hline
   (a) \;  {\it M2} & \times & & &  & & & & & & \times & \times \\
   (b) \;  {\it M5} & \times & \times & \times & \times & \times & & & & & \times & \\
   (c) \;  {\it M5} & & \times & \times &  &  & \times& \times & & & \times & \times\\
   (d) \;  {\it M5} & & \times & \times & \times & & & \times & \times & \times & & \\
 \hline
\end{array}
\ee
Given the structure of spacetime, the dual seven-form now has components ${\tilde {\cal F}}_{mnp {\bar q} {\bar r} {\bar s} a}$, ${\tilde {\cal F}}_{mnp {\bar q} {\bar s} ab}$ and ${\tilde {\cal F}}_{{\bar m} {\bar n} {\bar p} qs ab}$ and will thus be felt only by probes (c) and (d) in the table above. Neither (a) nor (b) sense the field strength and consequently have closed volume forms. Applying what is by now a familiar analysis, we obtain the following  conditions

\bit
\item
(a) Unlike earlier configurations ($n = 2,3$) where an analogous membrane probe implied  $d (H_1 H_2^2) = 0$, the paucity of transverse directions makes this conclusion a little more complicated. Since the membrane spans the entire transverse space, $\partial^{'} {\cal V}$ vanishes automatically, without necessarily implying $\partial^{'} (H_1 H_2^2) = 0$. In order for the volume form to be closed, the derivative $\partial {\cal V}$ must vanish as well, i.e
\be
\partial (H_1 H_2^2) \wedge dt \wedge dy^1 \wedge dy^2= 0
\label{hh4}
\ee
In the absence of any information about the transverse space derivative of $H_1 H^2_2$, the above condition is not sufficient reason to conclude that $H_1 = H_2^{-2}$, so we hold off on making that claim for now.
\item
(b) The volume-form of the fivebrane probe ${\cal V} = H_1 H_2 dt \wedge \Omega \wedge dy^a$ should be closed, leading to the condition
\be
d (H_1 H_2 \Omega) \wedge dt \wedge dy^2= 0
\label{omega5}
\ee
\item
(c) For the second fivebrane, ${\cal V}$ is given by $H_2^2 J \wedge J \wedge dy^a \wedge dy^b$. The derivative $\partial^{'} {\cal V}$ vanishes trivially, since all transverse indices are saturated by ${\cal V}$. Consequently, we can not obtain any information about $\partial^{'} (H_2^2 J \wedge J)$ from this probe. The term $\partial {\cal V}$, however, has the same index structure as ${\tilde {\cal F}}_{mnp {\bar q} {\bar s} ab}$ so we find
\be
\partial (H_2^2 J \wedge J) \wedge dy^a \wedge dy^b =  * \partial (H_1 J) \wedge dt
\ee
This can be used to conclude that $\partial \ln {\sqrt G} = \partial \ln H^{1/3}$
\item
(d) The last fivebrane probe has a volume-form ${\cal V} = J \wedge J \wedge J$. The derivative $\partial {\cal V}$ finds no counterpart in the expression for ${\tilde {\cal F}}$ and hence must vanish:
\be
\partial (J \wedge J \wedge J) = 0
\ee
The transverse space derivative $\partial^{'} {\cal V}$ has the same structure as ${\tilde {\cal F}}_{mnp {\bar q} {\bar r} {\bar s} a}$, so we set the two equal to each other. In other words,
\be
\partial^{'} (J \wedge J \wedge J) = * \partial^{'} (H_1 J)
\ee
This implies $\partial^{'} \ln {\sqrt G} = \frac{1}{3} \partial^{'} \ln H$, together with the conjugate expression,
and also
\be
\partial^{'} (H_1 J \wedge J \wedge J \wedge J) = 0
\label{cy4}
\ee
\eit
\no
We now have sufficient information to determine the relationship between $H_1$ and $H_2$. Using (\ref{omega5}) and the identity $\Omega \wedge \bar{\Omega} = J \wedge J \wedge J \wedge J$  we can make the following statement:
\be
\partial^{'} (H_1 H_2 \Omega \wedge H_1 H_2 \bar{\Omega}) = \partial^{'} (H_1^2 H_2^2 J \wedge J \wedge J \wedge J) = 0
\ee
Together with (\ref{cy4}), the above relation implies $\partial^{'} (H_1 H_2^2) = 0$. Putting this together with (\ref{hh4}) we can now conclude that $d(H_1 H_2^2) = 0$ so and the functions are indeed related to each other in the usual manner $H_1 = H_2^{-2}$.

To summarize, an M2-brane wrapped on a holomorphic 2-cycle in ${\cal M}_4$ gives rise to a background that can be described by the metric and four-form (\ref{mfs}) where the determinant of the metric $G_{m {\bar n}}$ is now $ G = H^{1/3}$. In addition, the manifold ${\cal M}_4$ must be such that
\bea
d (H^{-1/6}\Omega) = 0 \nonumber  \\
\partial (J \wedge J \wedge J) = 0  \nonumber  \\
\partial^{'} (H^{-1/3} J \wedge J \wedge J \wedge J) = 0 \nonumber
\eea
Einstein's equations are solved, provided
\be
\partial^{'} \partial^{'} (J \wedge J \wedge J) + 6 i \partial \bar{\partial} (H^{1/3} J \wedge J) = 0
\ee

\subsection{M2 on a Holomorphic Cycle in ${\cal M}_5$}

Consider, finally, the configuration in which the M2-brane wraps a holomorphic two-cycle in ${\cal M}_5$. Since this manifold spans the entire space, the metric is of the form
\be
ds_{11}^2 =  - H^{-2/3} dt^2 + 2 G_{m {\bar n}} dz^m dz^{\bar n}
\label{metric5}
\ee
We have chosen to denote the warp factor by $H^{- 2/3}$ instead of $H_1^2$, so that the expressions we obtain here are consistent with those for $n = 2, 3, 4$. Since there are no longer any transverse directions, the components of the field strength are simply
\bea
{\cal F}_{0mn{\bar p}} &=& \frac{i}{2} [\partial_{n} (H^{-1/3} G_{m {\bar p}}) - \partial_{m} (H^{-1/3} G_{n {\bar p}})] \nonumber \\
\label{f5}
\eea
and its complex conjugate ${\cal F}_{0 {\bar m} {\bar n} p}$. Consequently, ${\tilde {\cal F}}$ receives contributions only from ${\tilde {\cal F}}_{mnpq{\bar r}{\bar s}{\bar u}}$ and ${\tilde {\cal F}}_{{\bar m}{\bar n}{\bar p}{\bar q} r s u}$.

\no
The only possible supersymmetric probes of this geometry are the two fivebranes represented below:
\be
\begin{array}[h]{|c|c|cccccccccc|}
  \hline
   \; & 0 & 1 & 2 & 3 &
              4 & 5 & 6 & 7 & 8 & 9 & 10\\
  \hline
  {\bf M2} & \otimes & \otimes & &  & & & \otimes& & &  &  \\
  \hline
  (a) \; {\it M5} & \times & \times & \times & \times & \times & \times &  & & & & \\
  (b) \; {\it M5} & &  \times & \times &  \times & & & \times &  \times & \times & & \\

  \hline
\end{array}
\ee
These impose the following conditions on the background:
\bit
\item
(a)The first fivebrane has volume form ${\cal V} = H^{-1/3} dt \wedge \Omega$. Clearly, neither of the two components of $d{\cal V}$ will have the same structure as those of $\tilde {\cal F}$, so ${\cal V}$ must be closed, i.e
\be
\partial(H^{-1/3} \Omega) \wedge dt= 0, \;\;\;\;\;\;\; {\bar \partial} (H^{-1/3} \Omega) \wedge dt = 0
\label{m55}
\ee
\item
(b) This Euclidean fivebrane wraps the holomorphic six-cycle calibrated by $J \wedge J \wedge J$, so
$\partial {\cal V}$ has the same index structure as ${\cal F}_{mnpq {\bar r} {\bar s} {\bar u}}$. Setting these two quantities equal to each other gives the condition
\be
\partial(J \wedge J \wedge J) = * \partial (H^{-1/3} J)
\label{j5}
\ee
This implies that the determinant of the metric is given by $\partial \ln {\sqrt G} = \frac{2}{3} \partial \ln H$ and in addition
\be
\partial (H^{-1/3} J \wedge J \wedge J \wedge J) = 0
\ee
\eit
\no
To conclude, we find that the supersymmetric background of an M2-brane wrapped on a holomorphic 2-cycle in ${\cal M}_5$ is described by the metric (\ref{metric5}) and field strength (\ref{f5}) with the determinant now given by $\sqrt{G} = H^{2/3}$. The manifold ${\cal M}_5$ is constrained such that
\be
\partial (H^{-1/3} \Omega) = 0 \;\;\;\;\;\;\; \& \;\;\;\;\;\;\; \partial (H^{-1/3} J \wedge J \wedge J \wedge J) = 0
\ee
The above constitutes a supergravity solution when supplemented by the equation of motion for ${\cal F}$ which in this case is simply
$\partial \bar{\partial} (J \wedge J \wedge J) = 0$

\section{An M2-Brane on a SpelL 3-cycle}

Having exhausted all holomorphic cycles, the only way left in which to generate a BPS membrane background is to wrap a Euclidean M2-brane on the Special Lagrangian three-cycle in a CY 3-fold.

Suppose, for concreteness, that we wrap the M2-brane on the Special Lagrangian 3-cycle calibrated by ${\sf Re} \Omega$. Since the brane is entirely contained within the manifold ${\cal M}$ , the transverse directions should exhibit an $SO(4,1)$ rotational symmetry as reflected in the metric ansatz below $$ds^2 = H^{1/3} \eta_{a b} dy^a dy^b + G_{mn} dx^m dx^n$$ where $x^a, a = 0, \dots 4$ runs over the transverse coordinates while $x^m, m = 1 \dots 6$ span $\cal M$.

The M2-brane has volume form ${\cal V} = {\sf Re} \Omega$, so the field strength
\be
{\cal F} = d \; {\sf Re} \Omega
\ee
has components ${\cal F}_{amnp}$ and ${\cal F}_{mnpq}$. Consequently, the dual seven form ${\tilde{\cal F}} = \ast {\cal F}$ receives contributions from ${\tilde{\cal F}}_{abcdmnp}$ and ${\tilde{\cal F}}_{abcdemn}$.

Notice that we have not defined complex coordinates on ${\cal M}$. This is because now $d \Omega$ measures flux and hence can not vanish, therefore a complex structure does not exist on $\cal M$.

We now proceed to introduce supersymmetric probes and figure out which of these couple to the flux.
\be
\begin{array}[h]{|c|ccccc|cccccc|}
  \hline
   \; & 0 & 1 & 2 & 3 &
              4 & 5 & 6 & 7 & 8 & 9 & 10\\
              \hline
  {\bf M2} & &  & &  & & \otimes & \otimes & \otimes & &  &  \\
  \hline
   (a) \; {\it M2} & \times & & &  & & \times & & & \times &  &  \\
   (b) \; {\it M5} & \times & \times & &  & & \times & \times & & \times & \times &  \\
    (c) \; {\it M5} & \times & \times & \times &  & & \times & \times & &  &  & \times \\
   \hline
\end{array}
\ee

\bit
\item
(a) The membrane has worldvolume ${\cal V} = H^{1/6} dy^a \wedge J$. The fact that this is a closed form leads to the constraints
\be
 \partial (H^{1/6} J) = 0, \;\;\;\;\;\;\;\;\;\;\;\;  \partial^{'} (H^{1/6} J)
 \label{spell3}
\ee
where $\partial$ denotes a derivative along ${\cal M}$ whereas ${\partial}^{'}$ indicates a derivative in the transverse space.
\item
(b)
 The fivebrane with worldvolume ${\cal V} = H^{1/3} dy^a \wedge dy^b \wedge J \wedge J$ yields no new information, since setting $d{\cal V} = 0$ merely leads to conditions $ \partial (H^{1/3} J \wedge J) =0$ and $ \partial^{'} (H^{1/3} J \wedge J) =0$ which are trivial consequences of (\ref{spell3}) above.
\item
(c) The final fivebrane probe has volume ${\cal V} = H^{1/2} dy^a \wedge dy^b \wedge dy^c \wedge {{\sf Im} \Omega}$.
The component ${d {\cal V}}_{abcmnpq}$ of the seven-form $d {\cal V}$ has no counterpart in ${\tilde {\cal F}}$ and hence must vanish:
\be
\partial (H^{1/2} {{\sf Im} \Omega}) = 0
\label{fspell}
\ee
The transverse space derivative $\partial^{'} {\cal V}$ however, has the same index structure as ${\tilde {\cal F}}_{abcdmnp}$. Requiring these to be equal, we find
\be
\partial^{'} (H^{1/2} {{\sf Im} \Omega}) = \ast \partial^{'} {\sf Re} \Omega
\label{spel3}
\ee
This can be used to conclude $\sqrt {G} = H^{-1/2}$ and $\partial^{'} (H^{1/2} \Omega \wedge {\bar \Omega}) = 0$.
\eit

\no
 We find that the BPS background of an M2-brane wrapping a SpelL 3-cycle can be specified by the metric (\ref{spell3}) and field strength (\ref{fspell}) where the determinant of the metric is now given by $\sqrt {G} = H^{-1/2}$. In addition, the manifold ${\cal M}$ is required to obey the following constraints:
\bea
d(H^{1/6} J) &=& 0 \nonumber \\
\partial (H^{1/2} {{\sf Im} \Omega}) &=& 0 \nonumber \\
\partial^{'} (H^{1/2} \Omega \wedge {\bar \Omega}) &=& \partial^{'} (H^{1/2} J \wedge J \wedge J) = 0
\eea
The equation of motion for the field strength now takes the simple form $$d * d \; {\sf Re} \Omega = 0$$.

\section{Conclusions}

While various aspects of wrapped membrane configurations have been studied earlier \cite{membranes},\cite{amherst}, \cite{kastor} what we present here is the complete and most general form of the constraints on such geometries. Moreover, each constraint has a clear and obvious interpretation as a calibration in the background. As is clear from the examples we discussed, the probe approach drastically simplifies calculations and also -  perhaps more importantly -  paints an intuitive physical picture which makes the geometrical conditions on the backgrounds almost obvious, and the reasons behind them, transparent.

Having applied this method quite successfully to wrapped membrane backgrounds, the natural next step is to use it to study  supersymmetric backgrounds of wrapped M5-branes. This is a project that is currently under way. For the most part, the discussion presented here simply carries over; there is however, an added complication in that M5-branes can also have world-volume fields turned on. Incorporating these fields into the above framework is somewhat challenging and it will be interesting to see how it all works out.

\acknowledgments

As always, Ansar Fayyazuddin has my gratitude, for many useful comments on the draft, innumerable helpful conversations during the course of this work and very enjoyable collaborations on which this paper is based. I am grateful to Jerome Gauntlett for discussions, and for pointing out relevant references. It is a pleasure to acknowledge Cumrun Vafa' s warm hospitality and general encouragement.

\end{document}